\documentclass[prl,twocolumn,aps,showpacs,preprintnumbers,amsmath,amssymb,superscriptaddress]{revtex4}
\usepackage{graphicx}
\usepackage{dcolumn}
\usepackage{bm}
\usepackage{amsmath} 

\begin{document} 

\title{Predicting spinor condensate dynamics from simple principles}

%

\author{M. Moreno-Cardoner}
\affiliation{Departament d'Estructura i Constituents de la Mat\`eria,
Facultat de F\'isica, Universitat de Barcelona, E-08028 Barcelona, Spain}

\author{J. Mur-Petit}
\thanks{Present address: Clarendon Laboratory, University of Oxford, 
  Parks Road, Oxford OX1 3PU, U.~K.}
\email{jordi.mur@lac.u-psud.fr}
\affiliation{IFRAF and Laboratoire Aim\'e Cotton, CNRS \& Univ.\ Paris-Sud,
F-91405 Orsay, France}

\author{M. Guilleumas}
\affiliation{Departament d'Estructura i Constituents de la Mat\`eria,
Facultat de F\'isica, Universitat de Barcelona, E-08028 Barcelona, Spain}

\author{A. Polls}
\affiliation{Departament d'Estructura i Constituents de la Mat\`eria,
Facultat de F\'isica, Universitat de Barcelona, E-08028 Barcelona, Spain}

\author{A. Sanpera}
\affiliation{ICREA and Grup de F\'isica Te\`orica, Universitat Aut\`onoma de
Barcelona, E-08193 Bellaterra, Spain}

\author{M. Lewenstein}
\affiliation{ICREA and ICFO-Institut de Ci\`encies Fot\`oniques,
E-08034 Barcelona, Spain}

\date{\today}

\pacs{03.75.Kk,03.75.Lm,05.30.Jp,64.60.Cn}

\begin{abstract}
We study the spin dynamics of quasi-one-dimensional $F=1$ condensates both 
at zero and finite temperatures for arbitrary initial spin configurations.
The rich dynamical evolution exhibited by these non-linear systems
is explained by  surprisingly simple principles:
minimization of energy at zero temperature, and 
maximization of entropy at high temperature.
Our analytical results for the homogeneous case are corroborated 
by numerical simulations for confined condensates 
in a wide variety of initial conditions.
These predictions compare qualitatively well with 
recent experimental observations and can, therefore, 
serve as a guidance for on-going experiments.
\end{abstract}

\pacs{03.75.Mn, 03.75.Kk}

\maketitle

Spinor condensates realized by optically trapped ultra\-cold atoms 
in a hyperfine Zeeman manifold
allow to address 
a broad scope of problems, which are related to magnetic ordering. 
These include quantum phase transitions, exotic topological defects 
and spin domains either within a mean-field regime or 
within the strongly-correlated one \cite{lewen07}.
Already in the mean-field regime the dynamics of spinor condensates shows 
some intriguing features, with an apparent randomness  regarding the 
evolution from a given initial state towards a
steady state, 
accompanied by  formation of spin domains.
Even for the simplest case of an $F=1$ spinor condensate, 
the interplay between spin-spin interactions, non-linear terms   
and temperature effects rends the analysis of the dynamics rather complex. 
Previous studies of the spinor dynamics  
show that, at early stages of the evolution, 
there is a coherent population transfer between the different 
hyperfine coupled sublevels~\cite{Mur06,Pu99}.
Inclusion of temperature ($T$)  not only smears out the  
population transfer~\cite{Mur06}, 
but also leads to a  different distribution of population among
the different hyperfine levels.

In this contribution we analyze 
the dynamics of spinor $F=1$ condensates
both at zero and finite $T$  from a new perspective. 
As we shall show,
the complex dynamics displayed by these systems can be understood 
in terms of oscillations in phase space around a 
steady state. 
The  configuration of this  state can be approximately determined by 
analyzing the trajectories of constant 
energy in the homogeneous case. 
To a good approximation,
the populations that  characterize this state are rather close 
to those that minimize the energy associated to spin-spin interactions 
for a given magnetization.
In contrast, at finite $T$, 
the system is able to exchange energy with the 
thermal clouds. At large enough temperature, the populations of the steady state can be simply determined by maximizing the entropy of the homogeneous 
condensate.
This leads to a different long-time  configuration of the populations
than the one attained starting from the same initial conditions at $T=0$.
Our claims are supported by analytical results 
for an homogeneous condensate and by numerical investigations 
for trapped condensates. They  provide a good qualitative 
agreement with recent experimental data  on the dynamics 
of spinor condensates~\cite{nature,Chang04}. 
Furthermore, our 
results can be 
straightforwardly used to analyze 
 and predict experimental 
outcomes. 

In the mean-field approach, the spinor condensate is described by a vector order parameter
 $\Psi$ 
whose components $\psi_{m}$  correspond  to the order parameter of each magnetic sublevel 
$|F=1, m \rangle$
with $m=1,0,-1$. 
In absence of an external magnetic field and at $T=0$, 
the properties of the condensate are determined by the 
spin-dependent energy functional~\cite{Mur06,Ho98,machida}: 
\begin{align}
  {\cal E} = \, &   \int d^3r \left \{
    \psi^*_m \left(-\frac{\hbar^2}{2M}{\bm \nabla}^2
                   + V_{ext}\right) \psi_m \right. \nonumber\\
    &+ \left. \frac{c_0}{2} \psi^*_m \psi^*_{j} \psi_{j} \psi_m
    +\frac{c_2}{2}
      \psi^*_m\psi^*_{j} {\bf F}_{mk} \cdot
     {\bf F}_{jl} \psi_{l}\psi_k \right\} ,
\label{functional}
\end{align}
where repeated indices are assumed to be summed. 
The trapping potential is taken to be harmonic, 
axially-symmetric and spin independent.
The coefficients $c_0$ and $c_2$ describe the 
spin-independent and the spin-dependent contributions  
of the binary elastic collisions  of two identical spin-1 bosons.
They are expressed in terms of the $s$-wave scattering lengths $a_f$ 
of the combined symmetric channels of total spin $f=0,2$ as
$c_0=4 \pi \hbar^2 (a_0+2a_2)/(3M)$ and
$c_2=4 \pi \hbar^2 (a_2-a_0)/(3M)$, 
where $M$ is the atomic mass. ${\bf F}$ are the 
spin-1 matrices~\cite{Ho98,machida}. 
In this Letter, we focus on $^{87}$Rb 
spinor condensates~\cite{nature,Chang04,exper,Higbie05},
which have ferromagnetic character ($c_2<0$).

\begin{figure}[t]
\includegraphics[width=0.9\columnwidth,angle=0, clip=true]{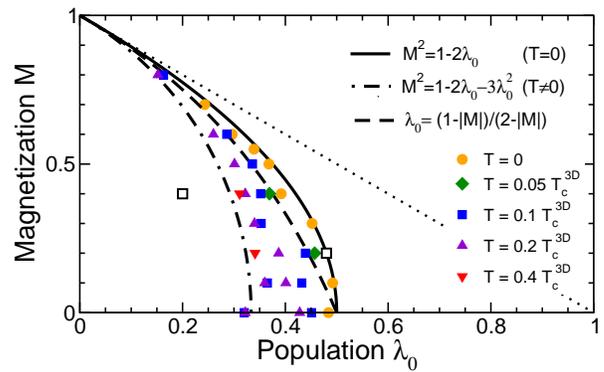}
\caption[]{(color online) 
  Equilibrium configurations in the $\lambda_0-{\cal M}$ plane for the 
homogeneous condensate at $T=0$ 
  [solid line, Eq.~(\ref{eq:lambda_eq})] 
  and at large  $T$ [dot-dashed line, Eq.~(\ref{eq:lambda_T})].
  The dashed line in between indicates $\lambda_0^{min}$, the geometrical center 
  of the closed orbit  with the same magnetization,
  while the dotted line bounds the region of
  physically allowed values for $(\lambda_0,{\cal M})$.
  Full symbols stand for the time averages of $\lambda_0(t)$  calculated 
  from the numerical time evolution for different initial configurations 
  at zero and finite $T$  for the inhomogeneous system;
  open squares indicate the initial conditions for the 
  simulations at ${\cal M} =0.2, 0.4$ respectively.
  Due to the symmetry of the equations, it suffices to present 
  the results for positive magnetizations.}
\label{fig1}
\end{figure}

According to $i \hbar \partial \psi_m/\partial t=
\delta {\cal E}/\delta \psi_m^*$, a set of  
coupled dynamical equations for the spin components are obtained:
\begin{subequations}
\begin{align}
i \hbar \, \partial \psi_{\pm1}/\partial t = \, &
 [{\cal H}_s  + c_2(n_{\pm1}+n_0- n_{\mp1})] \, \psi_{\pm1}  \nonumber \\
 &+ c_2 \, \psi_0^2 \psi^{*}_{\mp1} \,, 
 \\
i \hbar \, \partial \psi_0/\partial t = \, &
 [{\cal H}_s  + c_2(n_{1}+n_{-1})] \, \psi_{0} \nonumber \\
   &+ 2c_2 \psi_{1} \psi_0^* \psi_{-1} \,, 
%
\end{align}
\label{dyneqs}
\end{subequations}
with ${\cal H}_s=-\hbar^2/(2M)\, {\bm \nabla}^2 +V_{ext}+c_0 n$ being the
spin-independent part of the Hamiltonian.
The density of the $m$-th component is given by
$|\psi_m({\bf r})|^2$, while 
$n({\bf r})=\sum_m|\psi_m({\bf r})|^2$ 
is the total density normalized to the total number of atoms $N$.
The population of each hyperfine state 
is $N_m=\int d{\bf r} |\psi_m ({\bf r})|^2$. 
Defining the relative populations $\lambda_m=N_m/N$, 
the magnetization of the system is ${\cal M}=\lambda_1-\lambda_{-1}$.
The total number of atoms
and the magnetization are both conserved quantities~\cite{Mur06}.  
Then, the conditions  $\lambda_1+\lambda_0+\lambda_{-1}=1$ and 
${\cal M}=\lambda_1-\lambda_{-1}$, together with the positiveness 
of $\lambda_m$, restrict the allowed configurations of the system 
in the $(\lambda_0,{\cal M})$ phase space to the region below the dotted
line 
in Fig.~\ref{fig1}.

To understand the physics governing the spin dynamics, 
it is instructive to analyze first the homogeneous system at $T=0$. 
Translational invariance leads to the following ansatz:
$\psi_m = {\sqrt{n}} {\sqrt {\lambda_m}}\, e^{i \theta_m}$.
Notice that, since the system is homogeneous, 
the value of $n$  is not relevant to determine 
the relative populations 
that minimize the energy. 
Thus, the only relevant 
contributions to
the energy are those associated to $c_2$:
\begin{align}
  E_{c2}(\lambda_0,\cal{M},\theta) = \; & c_2 n^2 
  \big[ \lambda_0 (1 - \lambda_0)+ {\cal M}^2/2
  \nonumber\\ 
  &+ \lambda_0  
  {\sqrt{(1-\lambda_0)^2-{\cal M}^2}} \,
  \cos \theta \big] \,,
  \label{eq:ec2m}
\end{align}
where $\theta=2\theta_0-\theta_1-\theta_{-1}$. 
For a given value of ${\cal M}$, the ground state configuration 
is obtained by minimizing Eq.~(\ref{eq:ec2m}) with respect to 
$\lambda_0$ and $\theta$. For ferromagnetic systems, 
this yields $\theta=0$ and 
\begin{equation}
\lambda_0^{eq}=\frac{1}{2}(1-{\cal M}^2) \,,
\label{eq:lambda_eq}
\end{equation}
which defines the solid  line drawn in Fig.~\ref{fig1}.  
For instance, in the symmetric case 
 (${\cal M}=0$) the minimization leads to a ground state with 
$\lambda_0=0.5$ and $\theta=0$. 

%
We study now the dynamical evolution of the homogeneous system by solving 
numerically the set of Eqs.~(\ref{dyneqs}), 
after removal of the terms  $-\hbar^2/(2M)\, {\bm \nabla}^2 +V_{ext}$.
%
%
We consider three independent phases $\theta_m$ in the time evolution, 
although the energy 
depends only on the relative phase $\theta$ [Eq.~(\ref{eq:ec2m})].
As expected, we find that 
starting from an equilibrium configuration 
({\it e.g.} $\lambda_0=0.5$ and $\theta=0$ for ${\cal M}=0 $)
there is no evolution at all. 
However, if the initial configuration 
does not correspond to the equilibrium state,
then the system evolves exchanging populations
between spin components, but conserving energy and magnetization.

 Therefore, in the homogeneous system the dynamical evolution 
of the spin populations is 
characterized by trajectories of constant energy and magnetization 
in the $\lambda_0 - \theta$
plane, which are defined by the starting values of 
${\cal M}, \lambda_0$ and $\theta$.
In most cases, {\it i.e.} if the initial conditions 
are not very far from the equilibrium 
configuration, defined by $\lambda_0^{eq}$  and $\theta =0$, 
the trajectories in the $\lambda_0 - \theta$ plane are 
closed orbits (see also~\cite{You12005}). 
Then, the evolution of the populations --in particular of $\lambda_0$-- 
shows up as oscillations of a well defined frequency, 
around the geometrical center of the orbit. 
An analysis of the orbits defined by Eq.~(\ref{eq:ec2m}), 
shows that these centers lie always
between  $\lambda_0^{eq}$ and 
$\lambda_0^{min}= (1- |{\cal M}|)/(2-|{\cal M}|)$,
represented by the dashed curve in Fig.~\ref{fig1}.
With this information in mind we move now to the confined systems.

To this aim, we solve the equations of motion for 20000 atoms of $^{87}$Rb, 
in a highly elongated trap 
$\omega_{\perp}=2\pi\times 891$~Hz and $\omega_z=2\pi\times 21$~Hz, 
{\it i.e.} $\omega_{\perp} \gg \omega_z$, 
so that Eqs.~(\ref{dyneqs}) 
become quasi-one-dimensional. 
The corresponding scattering lengths are $a_0=101.8 a_B$ and $a_2=100.4 a_B$, 
and the coupling constants $c_0$ 
and $c_2$ are accordingly rescaled by a factor $1/(2 \pi a_\perp^2)$,
being $a_\perp=\sqrt{\hbar/ m \omega_\perp}$ the
transverse oscillator length ~\cite{foot1}.
The validity of this approximation for the case of a 
spin-1 atomic condensate has been
proved for a very wide range of conditions 
in Ref.~\cite{You12005}.
We take as initial wave functions $\psi_m(z,t=0)$ 
the wave function of the ground state of the scalar condensate, 
obtained from ${\cal H}_s$ with coupling constant 
$c_0/(2 \pi a_\perp^2)$, and normalized to their corresponding
initial populations $N_m$.
This corresponds to a Single Mode Approximation (SMA) ansatz for 
the initial conditions.  
SMA provides a good description of spin systems in 
certain conditions~\cite{Yi02,pendulum}.
However, in solving the time evolution equations we keep 
the full dependence of $\psi_m$
on position and therefore we go beyond the SMA. 
Actually, this is a necessary condition
to study the formation of spin domains.

The numerical solutions of the dynamical equations show oscillations 
of the integrated spin populations around a steady state,
which coincides with that in the homogeneous case [Eq.~(\ref{eq:lambda_eq})].
Indeed, if we start with a configuration lying 
 on the curve defined by $\lambda_0^{eq}$  with $\theta=0$,
the system does not evolve at all. However, any initial 
configuration with ($\lambda_0$,$\cal M$) on the curve, but $\theta \neq 0$,
yields  dynamical evolution of the spin populations around the
corresponding  steady configuration of the homogeneous case.

Our  results are summarized in Fig.~\ref{fig1}. 
The dots correspond to the time average value of $\lambda_0(t)$ 
obtained by  solving the equations of motion 
for different initial out-of-equilibrium values of 
$\cal M$, $\lambda_0$, and $\theta=0$.
 All  cases studied would correspond 
in the homogeneous system to closed orbits 
in the $\lambda_0-\theta$ plane. 
 The magnetization is constant throughout 
the evolution,
while $\lambda_0(t)$ presents slightly damped
oscillations 
around the same $\lambda_0$  as
in the homogeneous case, 
which is restricted to be in the narrow region limited by the solid 
and the dashed curves described above. 
Thus, the curve $\lambda_0^{eq}$ 
provides a  reliable 
estimation of the configuration of the steady state.
Notice that in all reported cases, the dots lie very close to the curve
$\lambda_0^{eq}$
as we have taken, in the selection  of the starting conditions
for ${\cal M} \ne 0$,  configurations and phases 
not far from the equilibrium conditions.

\begin{figure}[tb!]
\includegraphics[width=0.85\columnwidth,angle=0, clip=true]{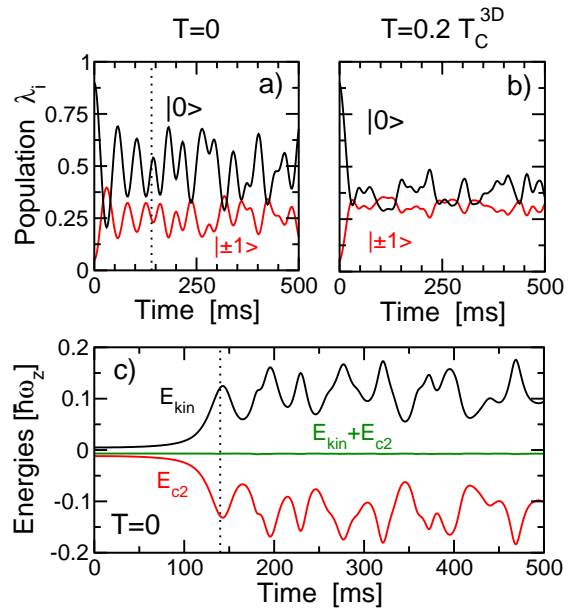}
\caption[]{(color online) 
  Spin dynamics for the initial configuration 
  $(\lambda_1,\lambda_0,\lambda_{-1})=(5\%,90\%,5\%)$ and $\theta=0$.
  Relative populations as a function of time 
  for (a) $T=0$ and (b) $T=0.2 T_c^{3D}$.  
   (c) Evolution of the kinetic (black) and $E_{c2}$ (red or dark grey) energies, 
  and their sum (green or gray) at $T=0$.
  The vertical lines are at $t_{dom}$.} 
\label{fig2}
\end{figure}

As an example of the dynamical evolution of the system  at $T=0$,  
we display in Fig.~\ref{fig2}(a) the spin dynamics 
for an initial symmetric spin configuration 
corresponding to $\lambda_1=\lambda_{-1}=5\%$ and $\theta=0$. 
The oscillations between $m=0$ and $m=\pm 1$ populations are not
regular and present a dynamical instability around $t_{dom}$,
when the condensate starts the multidomain formation process
\cite{Mur06,Robins01,Saito05,You2005}.
An estimate
of the time scale for the appearance of 
the instability is 
$t_{dom}\sim 2 \pi \hbar/(c_2 n_0)$ \cite{Saito05,You2005}, 
where $n_0$ is the  density at the center of the trap. 
In our simulation, $n_0 \sim 2 \times 10^{14}$ cm$^{-3}$, 
leading to  $t_{dom} \sim 140 $ ms, which is in agreement 
with the dynamics of the system. Around this time, the 
condensate starts the formation of  small dynamical 
spatial spin structures, also observed experimentally~\cite{nature,Higbie05},
while the relative populations oscillate around the equilibrium
state ($\lambda_0^{eq}=50\%$, $\lambda_{\pm 1}^{eq}=25\%$).
Due to this inhomogeneity of the phases and
density profiles~\cite{Mur06},
it is not possible to associate the dynamics to
a well-defined orbit  in the $\lambda_0-\theta$ plane.

It is interesting to note the formation of spin domains despite 
the total density profile remaining 
almost unchanged~\cite{Mur06,Saito05}.
Hence, spin domains are linked to the kinetic and spin-spin energy terms,
since all the other contributions to 
the energy functional depend only on the total density.
The evolution of the kinetic energy and $E_{c2}$ is shown in 
Fig.~\ref{fig2}(c) as a function of time for the first 500 ms.
Clearly the appearence of domains is linked to 
 a rapid variation
of kinetic and $E_{c2}$ energies. 
Moreover, since  the  system has no dissipation at $T=0$, 
the total energy  has to be conserved, and therefore, the energy difference 
between the starting and the equilibrium configurations  
provides an upper bound to the energy associated with  
the formation of spin domains. 
%
Actually this energy  is always rather small, 
as can be seen in the figure. 
In the case of Fig.~\ref{fig2}, the excess energy 
 amounts to 
0.28$\hbar \omega_z$ which represents a $0.3 \%$ of the total energy.

We incorporate now finite-temperature effects
by means of a Bogoliubov-de Gennes description of the thermal clouds.
Further we assume that each spin component has its own cloud~\cite{Tfinita}, 
and solve the corresponding time-dependent equations under
these conditions~\cite{Mur06}. 
With our choice of parameters, the condensation in an elongated 
trap occurs at  $k_B T_c^{3D} \approx 0.94 \hbar \omega_{ho} N^{1/3}$, where 
$\omega_{ho}= (\omega_{\perp }^2 \omega_z)^{1/3}$ \cite{Vandruten96}.
In Fig.~\ref{fig2}(b) we show for $T=0.2 T_c^{3D}$ the time evolution 
of the relative populations,  
starting from the same initial parameters as in the $T=0$ case.
It is worth  noticing that even if $T\gtrsim\hbar \omega_\perp$, 
the quasi-one-dimensional description of the thermal effects 
is still valid~\cite{Dettmer01}.

A simple comparison between Fig. 2(a) and 2(b)  
shows that at finite $T$ the oscillations in the populations 
are clearly damped and 
that the steady state configuration that is reached is 
different from the one at $T=0$.
%
In particular, in Fig.~\ref{fig2}(b) we observe that starting
from the same initial configuration $(5\%,90\%,5\%)$,
the long-time population distribution changes 
from $(25\%,50\%,25\%)$ at $T=0$  to 
a steady configuration with all $m$-states almost equally populated
(equipartition). 
 Moreover, even though the local magnetization is no longer
conserved as it was at $T=0$~\cite{Mur06},  
the total magnetization is still a conserved quantity along all the
time evolution.
In general, temperature effects reflect in a faster damping and 
in an earlier 
domain formation in comparison to $T=0$.
A way to understand this result is to assume that  
the  steady configuration is attained by 
maximizing the entropy, subjected to
the conservation of total population and magnetization:  
\begin{equation*}
  \bar S = - \sum_m \lambda_m \ln (\lambda_m) 
           - \Lambda \big(\sum_m m\lambda_m - {\cal M} \big) 
           - \mu \big(\sum_m\lambda_m -1 \big) ,
\end{equation*}
where we have introduced the Lagrange multipliers 
$\Lambda$ and $\mu$ associated 
with the magnetization and the normalization, respectively. 
Imposing that $ {\partial \bar S}/{ \partial \lambda_m}=0$, together with the 
normalization  and magnetization constraints, leads to the following 
condition for the steady state configurations:  
\begin{equation}
  {\cal M}^2 = 1 -2 \lambda_0 - 3 \lambda_0^2,
  \label{eq:lambda_T}
\end{equation}
which corresponds to the dot-dashed curve plotted in Fig.~\ref{fig1}. 
Notice that the curve thus obtained is independent of temperature. 
However, the hypothesis underlying its derivation  is appropriate
for a high temperature regime. 
Therefore, one  expects that this curve together with the one 
for the $T=0$ case [Eq.~(\ref{eq:lambda_eq})]  limit the zone of 
the configurations around which the system will oscillate 
depending on temperature.
In fact, the results of our numerical simulations for different temperatures and 
initial configurations, shown in Fig.~\ref{fig1},  are spread between the two
curves. We present results for $T/T_{c}^{3D}= 0.05$ (rhombi), 0.1 (squares), 0.2 (triangles)
and 0.4 (inverted triangles).  For a given initial configuration, when the temperature
increases, the steady state gets closer to the curve of maximum entropy. 
This is illustrated for
${\cal M}=0.2$ and 0.4. In both cases, the initial configuration is indicated 
by open squares.
Notice that for ${\cal M}=0.2$, the initial configuration is on the equilibrium curve ($T=0$)
and its evolution is a direct consequence of the temperature.
Also one observes a larger dependence of the  final steady state configuration 
on the initial conditions than at $T=0$.
In the figure, this is illustrated for ${\cal M}=0,$ and $0.1$ 
by the presence of  more than one symbol belonging to the same temperature 
that correspond to different initial conditions. 

Summarizing, we have shown that at zero temperature 
and in absence of external magnetic fields 
the spin populations
of $F=1$ condensates oscillate around 
a steady state which is well determined 
by analyzing the simpler homogeneous case. 
This steady state 
is close to the one 
described by minimizing the $E_{c2}$ energy 
at constant magnetization for the homogeneous case 
[Eq.~(\ref{eq:lambda_eq})]. 
In  confined systems,
spin domain formation is associated to the 
excess of spin-spin coupling energy, which 
converts into kinetic energy back and forward dynamically.

At finite $T$, the condensate exchanges energy with the thermal clouds, 
and the system evolves towards a different steady state configuration 
which depends also on the temperature. 
We have shown, however, that the configurations defined on the 
homogenous case by minimization of the energy and maximization of 
the entropy set an upper and lower bound on the accessible 
steady state configurations for confined spinor condensates at finite $T$. 
 These predictions can serve as a guide for the interpretation of 
on-going experiments on spinor condensates.
 

We thank K. Bongs and K. Sengstock for illuminating remarks.
We acknowledge financial support by  Spanish MEC 
Grants No.~FIS2005-03142/01414/01369/04627 and 
CONSOLIDER INGENIO2010 CDS2006-00019 ``QOIT'', 
by Generalitat de Catalunya Grants No.~2005SGR-00343/00185,
by the EU network ``Cold Molecules'' (HPRN-CT-2002-00290) 
and IST/FET/QIPC Integrated Project ``SCALA'', 
and by the ESF Programme ``QUDEDIS''.
Laboratoire Aim\'e Cotton is part of the F\'ed\'eration 
LUMAT and of IFRAF.

\end{document}